\documentclass{article}
\usepackage{waspaa23,amsmath,graphicx,url,times}
\usepackage{color}
\usepackage{multirow}
\usepackage{hyperref}
\usepackage{amssymb}
\usepackage{amsmath}

\usepackage{ifthen}
\title{Representation Learning for Audio Privacy Preservation using Source Separation and Robust Adversarial Learning}

\name{Diep Luong,$^{1}$
      Minh Tran,$^{1}$
      Shayan Gharib,$^{2}$
      Konstantinos Drossos,$^{3}$
      Tuomas Virtanen $^{1}$}
\address{$^1$ Tampere University, Tampere, Finland, \{diep.luong, minh.s.tran, tuomas.virtanen\}@tuni.fi \\
         $^2$ University of Helsinki, Helsinki, Finland, shayan.gharib@helsinki.fi\\
         $^3$ Nokia Tech, Espoo, Finland, konstantinos.drosos@nokia.com
}

\begin{document}

\ninept
\maketitle

\begin{sloppy}

\begin{abstract}
Privacy preservation has long been a concern in smart acoustic monitoring systems, where speech can be passively recorded along with a target signal in the system's operating environment. In this study, we propose the integration of two commonly used approaches in privacy preservation: source separation and adversarial representation learning. The proposed system learns the latent representation of audio recordings such that it prevents differentiating between speech and non-speech recordings. Initially, the source separation network filters out some of the privacy-sensitive data, and during the adversarial learning process, the system will learn privacy-preserving representation on the filtered signal. We demonstrate the effectiveness of our proposed method by comparing our method against systems without source separation, without adversarial learning, and without both. Overall, our results suggest that the proposed system can significantly improve speech privacy preservation compared to that of using source separation or adversarial learning solely while maintaining good performance in the acoustic monitoring task.
\end{abstract}


\section{Introduction}
\label{sec:intro}
The bloom of smart sensors has facilitated the process of data collection, enabling the integration of machine-learning approaches for various tasks. However, the transmission of data from local devices to remote servers for processing threatens user privacy: adversaries may access and exploit the data maliciously without the user's permission. One specific area that raises privacy concerns is acoustic monitoring, such as smart homes and smart cities, which targets recognizing pertinent sound events. While an acoustic monitoring system only targets non-speech sound events, most environments where such systems operate also contain speech. Speech signals contain personal information about speakers, including identity, gender, accent, as well as conversation content. Therefore, if attackers gain access to the speech signal, it could result in a dangerous privacy invasion for users.

\begin{figure*}[t!]
\centerline{\includegraphics[width=0.9\textwidth]{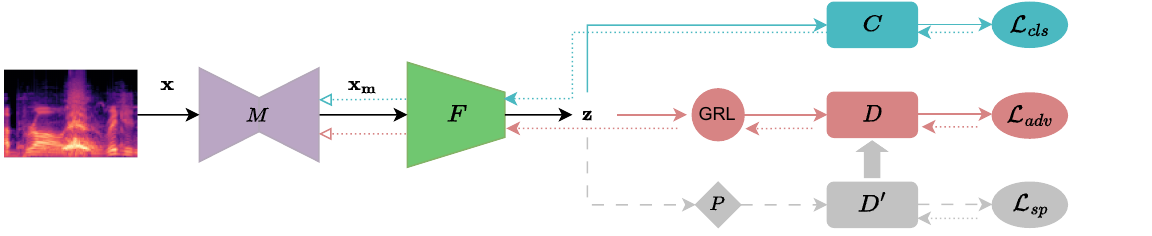}}
\caption{The schematic diagram of RDAL-M. $M$ is the source separation network, $F$ is the feature extractor, $C$ is the sound event classifier, $D$ is the speech discriminator on the adversarial branch, and $D'$ is the speech discriminator activated only after $P$ epochs. $\mathcal{L}$ represents different losses. The solid lines show the forward pass. The dashed line shows the forward pass to $D'$ only after every $P$ epochs. The dotted line shows the backpropagation from the losses to the corresponding weights. The dotted arrows with empty heads from $F$ to $M$ represent the backward pass in the \textit{learnable mask} approach. In the \textit{fixed mask} approach, there is no backpropagation from $F$ to $M$, and the parameters of the pre-trained $M$ are kept fixed during the training of RDAL-M.}
\label{fig:block_diagram}
\end{figure*}

Adversarial setups have emerged as a widely employed strategy for addressing privacy concerns in audio processing tasks. One of the first studies in the field is \cite{srivastava19_interspeech}, which proposed learning a representation that supports automatic speech recognition while being able to hide speaker information. The paper highlights the problem of recoverability of privacy information: performance of open set speaker recognition actually increases compared to when no privacy measure is taken. Noé et al. \cite{noe21_interspeech} used speakers' gender as a privacy criterion and proposed an adversarial setup between gender classification and automatic speaker verification. Perero-Codosero et al. \cite{PEREROCODOSERO2022101351} extended the adversarial setup to multiple privacy criteria: speaker ID, gender, and accent. The work demonstrates the superior privacy-preserving capabilities of multiple privacy branches in adversarial learning; however, the recovery of privacy information raised in \cite{srivastava19_interspeech} is yet to be addressed. The study in \cite{gharib2023adversarial} proposed robust discriminative adversarial learning (RDAL) as a solution for this problem by periodically optimizing a new speech discriminator during adversarial training. While RDAL achieved a substantial amount of reduction of information about speech activity, it did not reach the level of random guess which is the ultimate target, leaving room for additional techniques to improve the performance. 

In recent years, source separation has also gained attention as a privacy-preserving approach in audio processing. Cohen-Hadria et al. \cite{8918913} proposed a method of anonymizing the speech content and speaker's identity while having acoustic scenes preserved. Speech and non-speech signals are separated using a deep U-Net model, and the separated speech signal is then blurred. 
To deploy a light-weight privacy-preserving model for mobile systems, Xia and Jiang \cite{10.1145/3417313.3429383} proposed a sound source filtering algorithm called probabilistic template matching to generate a set of privacy-enhancing filters that remove privacy-sensitive signals using learned statistical ``templates'' of these sounds. The template is only built on a specific individual; thus, it cannot generalize to other people's speech signals. The proposed method in \cite{9929794} preserves privacy-sensitive speech content in cough events detection by integrating a source separation model to filter out speech content before performing cough detection. 
In speech-overlapping conditions, speech separation is used to extract only speech from a target user, preserving the privacy of non-target speakers \cite{aloufi21_interspeech} \cite{9369854}. Although source separation can be used for source-wise processing to preserve audio privacy, it can not completely remove the presence of privacy-sensitive information in the audio while posing a degradation in the performance of the utility task.  

To address the setbacks of adversarial and source separation in speech privacy preservation stated above, we investigate the integration of both approaches into a privacy-preserving system. In this study, we further enhance privacy preservation by incorporating a source separation network into the existing RDAL system in \cite{gharib2023adversarial}, thereby improving the efficiency of speech information removal. We show that the proposed method enables a better performance compared to either of the separate components.

The paper is structured as follows: a masking-based approach to adversarial representation learning is described in Section \ref{sec:method}, the system's performance in preventing speech presence detection from audio representation is evaluated in Section \ref{sec:evaluation}, and the study concludes with a summary of findings.

\section{Method}
\label{sec:method}
\subsection{Problem setup}
From the input audio features $\mathbf{x}$, our target is to derive a latent representation $\mathbf{z}$ which allows identifying the sound event class $\mathbf{y}$ while minimizing the exposure of privacy-sensitive attribute $\mathbf{s}$ of the speech signal. In this study, $\mathbf{x}$ is the short-time Fourier transform (STFT) magnitude spectrogram of the input audio signal, $\mathbf{y}$ represents sound event labels using one-hot encoding, and $\mathbf{s}$ is the binary classification label of speech and non-speech.
As the proposed method of privacy-preserving is generic, speech attributes can also be added to $\mathbf{s}$ as the privacy measures in the learning process.

\subsection{Robust discriminative adversarial learning with masking}
The proposed robust discriminative adversarial learning with masking (RDAL-M), illustrated in Fig.~\ref{fig:block_diagram}, includes a source separation network $M$, a feature extractor $F$ for representation extraction, a sound event classifier $C$ to identify the target sound event in the audio recording, and a speech discriminator $D$ which classify the sample into speech and non-speech. We integrate the source separation network with the adversarial system in two approaches: \textit{fixed mask} and \textit{learnable mask}. In the \textit{fixed mask} approach, $M$ is pre-trained for speech removal purposes and kept fixed during the adversarial training. In the \textit{learnable mask}, the training of $M$ is within the adversarial training process. 

For the training of RDAL-M, the dataset $\mathbf{X}$ contains $N$ training samples $\mathbf{x}$ with sound event labels $\mathbf{y}$, and speech label $\mathbf{s}$, i.e. $\mathbf{X} = \{(\mathbf{x}_i, \mathbf{y}_i, \mathbf{s}_i)^N_{i=1}\}$. The pre-training process of $M$ in \textit{fixed mask} utilizes $\mathbf{X^s}$, a subset of $\mathbf{X}$ which consists of $N_s$ samples that contain both speech and sound events.

The source separation network $M$ learns a mask to estimate the magnitude spectrogram of sound events only. Specifically, $M$ outputs a continuous mask in the range of $[0,1]$ denoted as $f_{\theta_M}(\mathbf{x})$. The separated spectrogram $\mathbf{x_m}$ is estimated by element-wise multiplying the input spectrogram by the mask. In the \textit{fixed mask} approach, given $\mathbf{x^s} \in \mathbf{X^s}$, $M$ is optimized by minimizing the mean absolute error between the estimated spectrogram $\mathbf{x^s_m} = \mathbf{x^s} \otimes f_{\theta_M}(\mathbf{x^s})$ and the target spectrogram of isolated sound events $\mathbf{x^s_{e}}$ within $\mathbf{X^s}$ as

\begin{equation}
    \label{eq:L_mask}
    \min_{M}\mathcal{L}_{mask} = \frac{1}{N_s} \sum_{i=1}^{N_s} \lVert \mathbf{x}^\mathbf{s}_{\mathbf{m}_i} - \mathbf{x}^\mathbf{s}_{\mathbf{e}_i}\rVert.
\end{equation}

The latent representation $\mathbf{z}=F(\text{log-mel}(\mathbf{x_m}))$ is extracted from the log-mel spectrogram of the filtered signal. To achieve good sound event detection (SED) capability, the feature extractor $F$ and sound event classifier $C$ are optimized together over $N$ training samples with the objective function
\begin{equation}
    \label{eq:l_cls}
    \min_{F, C}\mathcal{L}_{cls} = - \frac{1}{N} \sum_{i=1}^{N} {\mathbf{y}}_{i} \log(\hat{\mathbf{y}}_{i}),
\end{equation}
which is the cross-entropy loss between the true sound events $\mathbf{y}$ and the predicted sound events $\mathbf{\hat{y}}=C(\mathbf{z})$.

To prevent speech activity detection (SAD) in the audio representations, the optimization process also follows a minimax criterion between $F$ and a speech discriminator $D$:
\begin{equation}
\label{eq:L_adv}
\max_{F}\min_{D} \mathcal{L}_{adv} = -\frac{1}{N} \sum_{i=1}^{N} \mathbf{s}_i \log \mathbf{\hat{s}}_i + (1-\mathbf{s}_i) \log (1-\mathbf{\hat{s}}_i).
\end{equation}

Here, binary cross entropy is used as the loss between the true speech label $\mathbf{s}$ and the corresponding predicted output  $\mathbf{\hat{s}}=D(\mathbf{z})$. The goal is for $D$ to be optimized for the classification of speech samples, while $F$ tries to minimize such capabilities. With the inspiration of \cite{pmlr-v37-ganin15}, a gradient reversal layer (GRL) module is placed between $F$ and $D$. During backpropagation, the GRL module multiplies the gradient of $\mathcal{L}_{adv}$ with a negative number before updating the weights of $F$:

\begin{equation}
\label{eq:param_update_F}
\theta_F \longleftarrow \theta_F - \mu \left(\frac{\partial \mathcal{L}_{cls}}{\partial \theta_F} - \lambda \frac{\partial \mathcal{L}_{adv}}{\partial \theta_F}\right).
\end{equation}

This optimization process, however, does not guarantee that the sensitive speech information is irrecoverable. Therefore, adapting from \cite{gharib2023adversarial}, a new speech discriminator $D'$ is presented, which is optimized once after every $P$ epochs of adversarial learning with the objective function where $\mathbf{\hat{s}'} = D'(\mathbf{z})$
\begin{equation}
\label{eq:verify_loss}
\min_{D'}\mathcal{L}_{sp} = -\frac{1}{N} \sum_{i=1}^{N} \mathbf{s}_{i} \log \mathbf{\hat{s}'}_i
+ (1 - \mathbf{s}_{i}) \log (1-\mathbf{\hat{s}'}_i).
\end{equation}

After being optimized, the weights of $D'$ are shared with $D$ to continue the adversarial process. The supervised optimization of $D'$ ensures that the performance of the adversarial discriminator always stays at the maximum level and the adversarial process occurs efficiently.

In the \textit{learnable mask} approach, $M$ is optimized in the adversarial training process, and the parameters of $M$ are updated using the gradients from $\mathcal{L}_{cls}$ and $\mathcal{L}_{adv}$:
\begin{equation}
\label{eqp:param_update_M}
\theta_M \longleftarrow \theta_M - \mu \left(\frac{\partial \mathcal{L}_{cls}}{\partial \theta_M} - \lambda \frac{\partial \mathcal{L}_{adv}}{\partial \theta_M}\right).
\end{equation}

\section{Evaluation}
\label{sec:evaluation}
To assess the effect of source separation and adversarial learning, we implement an ablation study with the following components: the source separation network $M$, the feature extractor $F$, the sound event classifier $C$, and the speech discriminator $D$ on the adversarial branch.  Specifically, we compare the performance of RDAL-M in preserving speech privacy against the baseline, RDAL, and the masking method.

In the baseline, no privacy measures are taken, which means that $F$ and $C$ are trained jointly in a supervised manner for the SED task. RDAL preserves audio privacy by having an adversarial branch with the speech discriminator $D$ in the feature learning process \cite{gharib2023adversarial}. The masking method uses the pre-trained source separation network $M$ to remove speech, and thereafter, $F$ and $C$ are optimized in a supervised manner while keeping $M$ fixed. The output from $M$ is either \textit{continuous mask} in the interval $[0,1]$ or \textit{binary mask} with a threshold of 0.4: mask values smaller than 0.4 are set to 0, and the remaining ones are set to 1. The results for both approaches are reported to investigate the speech privacy preserving efficiency of the masking method with supervised feature learning.

\subsection{Dataset}
This study employs one-second audio recordings of either the mixture of speech and sound events or the sound events only. The dataset used in this study is similar to that from \cite{gharib2023adversarial}. 

The sound event data originate from FSD50K \cite{10.1109/TASLP.2021.3133208}. Dog Barking, Glass Breaking, and Gun Shot are selected as the target sound event classes.
The most energetic one-second segment is extracted from each audio sample and normalized by subtracting its mean and dividing by the standard deviation.

The speech contents are provided by LibriSpeech corpus \cite{7178964}. The speech signals are resampled from 16 kHz to 44.1 kHz to match the sampling rate of the sound events. The most energetic one-second segment is extracted from each sample and normalized following the same process discussed above. The extracted segments are attenuated by 5 dB, and the mixtures are created by adding the speech content from Librispeech train-clean-100 and dev-clean and the segments of sound event recordings from FSD50K development and evaluation sets together. To create a balanced dataset between speech and non-speech samples, the number of one-second segments extracted from speech recordings in LibriSpeech is half the number of sound event segments.

\begin{table}[t]
    \caption{Number of samples for each sound event and the number of samples containing speech in each split}
    \begin{center}
    \begin{tabular}{|c|c|c|c|c|}
    \hline
    \multicolumn{2}{|c|}{\textbf{}} & \textbf{Train} & \textbf{Validation} & \textbf{Test} \\
    \hline
    \multirow{2}{*}{\text{Sound}} & \text{Dog barking} & 374 & 40 & 122 \\
    \multirow{2}{*}{\text{event}} & \text{Glass breaking} & 374 & 40 & 96 \\
    & \text{Gun shot} & 314 & 34 & 134 \\
    \hline
    \multicolumn{2}{|c|}{\text{Speech}} & 531 & 57 & 176 \\
    \hline
    \end{tabular}
    \label{table-data-samples}
    \end{center}
\end{table}

\begin{table*}[t!]
    \caption{Results table of the ablation study with source separation and adversarial learning}
    \begin{center}
    \begin{tabular}{|c|c|c|c|c|c|c|}
    \hline
    \multirow{4}{*}{\textbf{Method}} & \multicolumn{2}{|c|}{\textbf{no masking}} & \multicolumn{4}{|c|}{\textbf{with masking}} \\
    \cline{2-7}
    
    \textbf{} & \textbf{supervised} & \textbf{adversarial} & \textbf{\textit{continuous mask}} & \textbf{\textit{binary mask}} & \textbf{\textit{fixed mask}} & \textbf{\textit{learnable mask}}\\
    \cline{4-7}
    
    \textbf{} & \textbf{feature learning} & \textbf{feature learning (RDAL)} & \multicolumn{2}{|c|}{\textbf{supervised}} & \multicolumn{2}{|c|}{\textbf{adverarial feature}} \\

    \textbf{} & \textbf{} & \textbf{} & \multicolumn{2}{|c|}{\textbf{feature learning}} & \multicolumn{2}{|c|}{\textbf{learning (RDAL-M)}} \\
    \hline
    
    \text{SED\textsubscript{accuracy}} & $0.84\pm0.01$ & $0.84\pm0.02$ & $0.84\pm0.01$ & $0.84\pm0.01$ & $\mathbf{0.86\pm0.01}$ & $0.84\pm0.02$ \\
    
    \text{SAD\textsubscript{accuracy}} & $0.79\pm0.02$ & $0.57\pm0.04$ & $0.57\pm0.04$ & $0.75\pm0.01$ & $\mathbf{0.50\pm0.03}$ & $0.55\pm0.04$ \\
    
    \text{AUC score} & $0.88\pm0.02$ & $0.60\pm0.05$ & $0.59\pm0.05$ & $0.83\pm0.01$ & $\mathbf{0.50\pm0.03}$ & $0.58\pm0.06$ \\
    \hline
    
    \end{tabular}
    \label{table-results}
    \end{center}
\end{table*}

Two disjoint sets, development and test, are obtained after creating the mixtures. For the training of the RDAL-M system, 90\% of the development set is chosen randomly to be dedicated to the training set, and the remaining 10\% goes to the validation set. For training the source separation network $M$ of RDAL-M \textit{fixed mask} approach and the masking method, we utilized the mixtures from the training set as the model input and the corresponding sound event segments before merging as the ground truth. Table~\ref{table-data-samples} shows the number of samples of each sound event and the number of samples containing speech within each split. For each sound event in each split, we ensure that half of the samples contain speech. 

\subsection{Model architecture}
The source separation network $M$ takes the STFT magnitude as the low-level feature input. The STFT is computed using a Hamming window with a window size of 32 ms and a hop length of 10 ms. These parameters are chosen based on \cite{kong20} and scaled based on the sampling frequency of the audio recordings. The input of our feature extractor $F$ is the log-mel spectrogram calculated on the output of the source separation network $M$ with 64 mel filters.

The source separation network $M$ utilizes the U-Net structure, which was originally designed for cell segmentation \cite {10.1007/978-3-319-24574-4_28}. 
The study in \cite{andreas_jansson_2017_1414934} successfully applied this structure to separate singing voice from non-vocal elements. Our source separation network $M$ follows the architecture from \cite{andreas_jansson_2017_1414934}. 
Encoder includes six 2D convolutional blocks, each consisting of a convolutional layer with kernel size $5\times5$ and a stride of 2 followed by Batch Normalization and LeakyReLU. The first block consists of 16 convolutional filters and it doubles at each block. The decoder has the same number of blocks in which convolutional layers are replaced by transposed convolutions and LeakyReLU by ReLU. It halves the number of filters at each block until it has 16 feature maps. The last layer outputs the mask and uses Sigmoid as the activation function.

Similar to \cite{gharib2023adversarial}, the architecture of the feature extractor $F$ is adapted from the CNN6 architecture in \cite{kong20}. $F$ includes 4 convolutional blocks; each block consists of a 2D convolutional layer with the kernel size $3\times3$ followed by ReLU, and Batch Normalization. The number of filters in the convolutional blocks is 64, 128, 256, and 512, respectively. Max pooling of size $2\times2$ is applied after the first 3 convolutional blocks, and global max pooling is applied to the output of the last convolutional layer to transform 2D data into a 1D feature vector. Finally, a fully connected layer reduces the dimension of the 1D representation from 512 elements to 64 elements. $C$ consists of one linear layer with softmax activation. $D$ includes 4 linear layers. The first 3 layers have output dimensions of 48, 32, and 16, and each layer is followed by LeakyReLU. Sigmoid is applied on the output layer.

\subsection{Training}
In the \textit{fixed mask} approach of RDAL-M and the masking method, the source separation network $M$ is trained with an Adam optimizer with a learning rate of 0.001. The training for $M$ has the patience of 20 epochs on the validation loss, and the parameters of the trained source separation network $M$ are kept fixed during the adversarial training process. 

We use the same training specifications as in \cite{gharib2023adversarial} for RDAL-M. $\lambda$ is set to 0 for the first 30 epochs to ensure that the speech classifier $D$ is adequately trained to classify between speech and non-speech samples from the beginning of the adversarial training.

\begin{figure}[t!]
\centerline{\includegraphics[width=0.9\columnwidth]{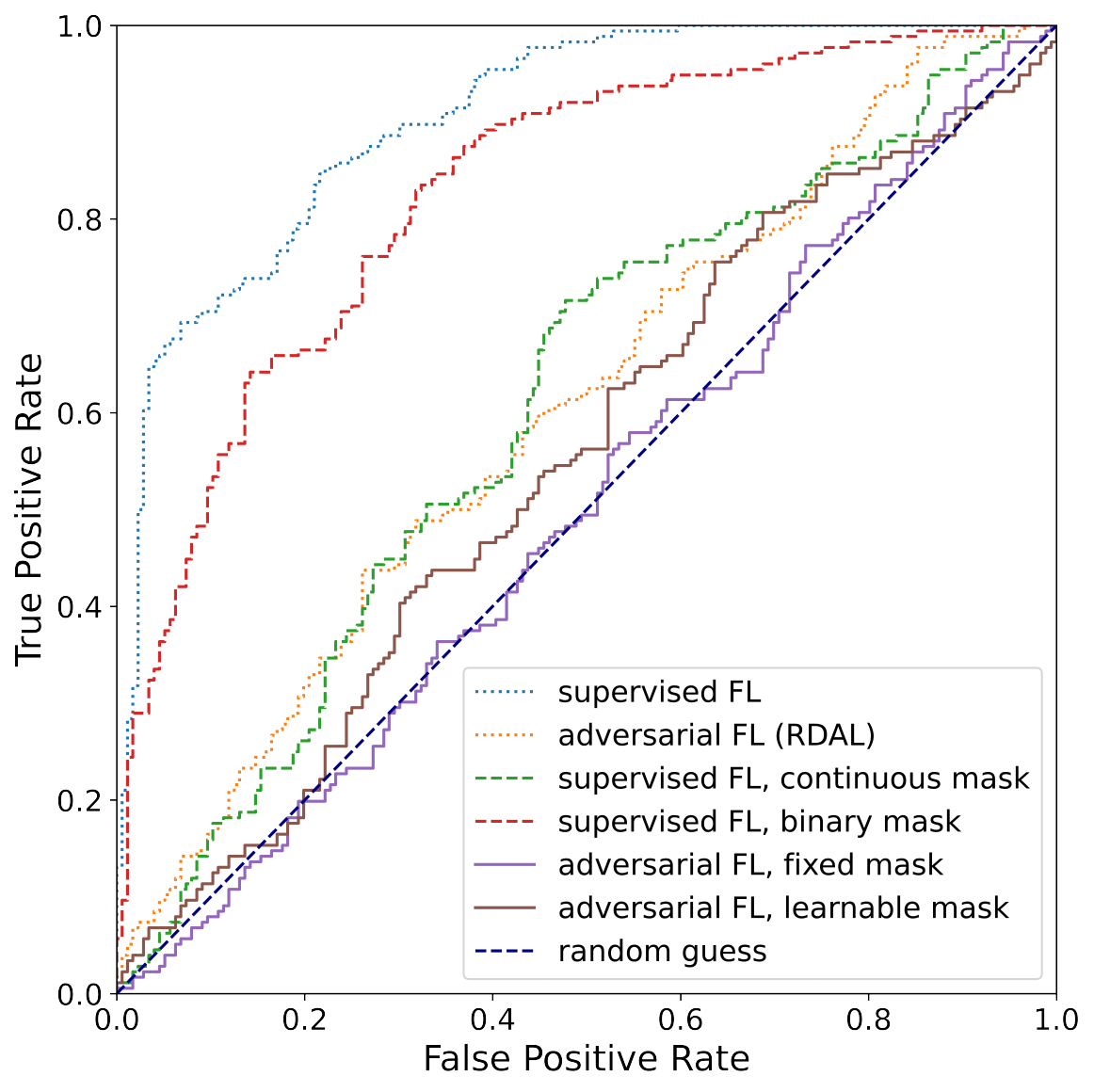}}
\caption{ROC curves of different methods discussed in Tables~\ref{table-results}. FL stands for feature learning.}
\label{fig:roc_curve}
\end{figure}

\subsection{Results}
Table~\ref{table-results} presents the SED accuracy, SAD accuracy, and AUC score on the test set of the baseline, RDAL, masking method, and RDAL-M. We take the accuracy and the AUC score in classifying speech and non-speech signals as the metrics to evaluate the performance in hiding privacy-sensitive content. A new speech classifier is trained over the representations of the trained feature extractor $F$, simulating the attacks where privacy intruders try to extract information from speech activities encoded in the latent space. The reported results are obtained as the mean and standard deviation of 10 separate runs for each approach. The SED and SAD accuracy of baseline and RDAL are taken from \cite{gharib2023adversarial}, and we calculate the AUC score ourselves.

From Table~\ref{table-results}, we can have the following observations. Firstly, the performance of RDAL-M on SED tasks is not compromised by the application of the source separation network. In terms of privacy preservation, the \textit{fixed mask} approach of RDAL-M demonstrates a significantly better gain in privacy considering the SAD accuracy and the AUC score compared to RDAL and the masking method. In detail, RDAL-M obtains both the SAD accuracy and the AUC score of 0.50 for the \textit{fixed mask} approach, which equals the random guess rate for a binary classification task. The ROC curves presented in Fig.~\ref{fig:roc_curve} demonstrate RDAL-M's superior privacy preservation capability compared to RDAL and the masking method. Notably, the efficacy of the binary classification task for RDAL-M \textit{fixed mask} is observed to be proximal to a random guess. We also measure the source separation capability of $M$ in different approaches using signal-to-distortion ratio (SDR), which is a common metric for source separation evaluation. The SDR of the pre-trained source separation network $M$ in \textit{fixed mask} is $11.72\pm0.05$, and the SDR of $M$ in \textit{learnable mask}, following its training in an adversarial process, is $3.45\pm1.03$. The results indicate that the \textit{learnable mask} provides slightly better privacy preservation compared to RDAL and the masking method. This suggests that $M$ can effectively identify and filter out privacy-sensitive information, and the filtered information may not necessarily be speech. Lastly, the SAD accuracy and the AUC score in the masking method show that source separation alone cannot fully hide the presence of the speech signal in an audio recording although the source separation network $M$ from the masking method has relatively high SDRs of $11.64\pm0.18$ for \textit{continuous mask} and $11.13\pm0.16$ for \textit{binary mask}.

\section{Conclusion}
Acoustic monitoring systems are prone to privacy leakage of users' information. This study defines the problem of privacy violation as detecting speech activity from the latent representation of audio recordings. Our proposed method, RDAL-M, aims to learn the latent representations of audio recordings in a way that prevents the differentiation between speech and non-speech recordings. RDAL-M utilizes both source separation and adversarial learning: a source separation network partially removes privacy-sensitive signals and an adversarial learning setup between a feature extractor and a speech classifier to preserve audio privacy. Specifically, the feature extractor learns a privacy-preserving representation, while the speech classifier tries to distinguish the speech presence from the representation. Results from our evaluation demonstrate the effectiveness of this method in obscuring speech presence within audio recordings while significantly preserving the performance of the utility task, thereby mitigating privacy violations resulting from passive speech recording. The results also indicate that RDAL-M performs equally to a random guess in a binary classification task on the speech presence, hence offering better privacy preservation than using only source separation or adversarial learning techniques alone.

\bibliographystyle{IEEEtran}
\bibliography{refs23}
%
%
%
%
%
%
%
%
%

\end{sloppy}
\end{document}